\documentclass[aps,prl,twocolumn,groupedaddress]{revtex4-1}

\usepackage{graphicx}
\usepackage{color}

\newcommand{\mnras}{MNRAS}
\newcommand{\ssr}{Space Science Reviews}
\newcommand{\apss}{Astrophysics \& Space Science}
\newcommand{\apjl}{ApJ}
\newcommand{\procspie}{Proc.~SPIE}

\begin{document}

\title{Inductive spikes in the Crab Nebula --- a theory of gamma-ray flares}
\author{John~G.~Kirk}
\email[]{john.kirk@mpi-hd.mpg.de}
\affiliation{Max-Planck-Institut f\"ur Kernphysik, Postfach 10 39 80, 69029 Heidelberg, Germany}
\author{Gwenael Giacinti}
\email[]{gwenael.giacinti@mpi-hd.mpg.de}
\affiliation{Max-Planck-Institut f\"ur Kernphysik, Postfach 10 39 80, 69029 Heidelberg, Germany}

\date{\today}
 
\begin{abstract}
  We show that the mysterious, rapidly variable emission at
  $\sim400\,$MeV observed from the Crab Nebula by the AGILE and Fermi
  experiments could be the result of a sudden drop in the mass-loading of
  the pulsar wind. The current required to maintain wave activity in
  the wind is then carried by very few particles of high Lorentz
  factor. On impacting the Nebula, these particles produce a tightly
  beamed, high luminosity burst of hard gamma-rays, similar to those
  observed. This implies (i) the emission is synchrotron
  radiation in the toroidal field of the Nebula, and, therefore,
  linearly polarized and (ii) this mechanism potentially
  contributes to the gamma-ray emission from other powerful pulsars,
  such as the Magellanic Cloud objects J0537$-$6910 and B0540$-$69.
\end{abstract}

\pacs{95.30.Jx, 95.30.Qd, 95.83.Sz, 97.60.Gb}
\maketitle

The detection of powerful gamma-ray flares from the Crab Nebula by the
AGILE satellite and the Large Area Telescope on the Fermi satellite
\cite{Agileflares11,Fermiflares11,buehleretal12} has provided
theorists with three major puzzles: How are particles able to emit
synchrotron radiation well above the $\sim100\,\textrm{MeV}$
astrophysical \lq\lq upper-limit\rq\rq~\cite{guilbertetal83}? What is the
geometry and location of the source, given that it varies on a
timescale of hours, whereas the Nebula has a light-crossing time of
months? By which mechanism can such a small source achieve a power
only one order of magnitude less than that of the entire Nebula?  Many
theories have addressed these issues 
(for a review see \cite{buehlerblandford14}),
but none has yet achieved general acceptance. In
this {\em Letter} a novel theory is proposed, based on the properties of
relativistic winds that are dominated by Poynting flux: the frequency,
variability and power of the flares emerge as natural consequences of
a sharp reduction of the supply of electron-positron pairs to the wind
of the Crab pulsar, an effect closely analogous to the voltage spikes
generated when the current in an inductive electrical circuit is
interrupted. 
Although the underlying cause remains a mystery, 
such interruptions are not implausible, 
since the electromagnetic cascades responsible for creating
the pairs are thought to be highly 
erratic~\cite{ceruttibeloborodov17}, 
a conclusion supported by
strong pulse-to-pulse fluctuations in the radio emission, that
presumably originates in this region~\cite{hankinsetal16}.  The new 
theory predicts the polarization properties of the flares, which may
be measurable in the near future \cite{tatischeffetal16}, and suggests that
similar emission may be detectable from other pulsar wind nebulae.

The Crab Nebula is powered by an electron-positron wind that is energetically dominated 
by electromagnetic fields. These oscillate at the pulsar
period $P$ and have a finite phase-averaged (DC) component~\cite{reesgunn74}, the 
necessary currents being carried by the pairs. Such winds propagate
radially \cite{kirketal09} up to a {\em termination shock} at 
radius $r=r_{\rm t.s.}$, where the ram pressure balances that of the surroundings. 
Because the particle density drops off as $1/r^2$, non-MHD effects become important
at sufficiently large radius. These can lead to the
conversion of the oscillations into electromagnetic waves
\cite{usov75,arkakirk12}, for which there is no observational evidence, or to 
damping of the oscillations, accompanied by radial acceleration of the plasma.  

This
latter process was originally modeled as magnetic
reconnection in a \lq\lq striped wind\rq\rq --- one with 
oppositely directed bands of toroidal magnetic field separated by hot
current sheets 
\cite{coroniti90,lyubarskykirk01}. Dissipation 
in the current sheets releases the magnetic tension, leading 
to radial acceleration of the
flow. However, in the case of the Crab, 
complete dissipation of the wave energy occurs only for a relatively high pair density
\cite{kirkskjaeraasen03}, which implies a terminal
Lorentz factor $<10^4$. An additional acceleration mechanism is then needed for the flares~\cite{zrake16}. 
Conversely, lower density flows reach higher bulk Lorentz factors, 
even though they do not achieve complete dissipation.
However, in the Crab, hot plasma cannot be confined in the sheets
up to the termination shock \cite{lyubarskykirk01}, which  
invalidates this model, and leaves the ultimate
fate of the waves uncertain. Here, we present a 
solution to this problem: assuming that the
pulsar wind is launched as a
mildly supersonic MHD flow with embedded magnetic fluctuations, we
demonstrate that inductive acceleration converts $10\,\%$ of
the power into kinetic energy. When the supply of electron-positron
pairs is severely limited, the 
few particles that are present 
achieve very high Lorentz factors.

Dissipation is not essential for acceleration
\cite{spruitdrenkhahn04}, as was shown using a static,
sinusoidal, magnetic shear (a \lq\lq sheet-pinch\rq\rq) as a wave
model~\cite{kirkmochol11}.  This wave contains neither a current sheet
nor hot plasma.  The transverse magnetic field has constant magnitude
and rotates at a uniform rate as a function of phase. The current is
carried by cold electron and positron fluids which, in the co-moving
frame, flow along the magnetic field. Provided the transverse flow
speed $v_\bot$, remains small, the wave propagates at constant speed,
as expected in MHD. But towards larger radius the drop in density
causes $v_\bot$ to increase to maintain the wave currents.  When
$v_\bot\approx c$, particle inertia causes a small misalignment of
current and field, leading to a net radial acceleration. Because
resistive dissipation is absent, and the underlying cause is the
maintenance of a current flow, this mechanism is appropriately
described as {\em inductive} acceleration.

Reference~\cite{kirkmochol11} identified
the location of the inductive acceleration zone, and gave an
approximate solution in which the Lorentz factor $\gamma$ of the
plasma is proportional to $r$.  This result is
not immediately applicable to a pulsar wind, because it does 
not allow for a DC component.  However, it is
straightforward to generalize the model to that of a 
striped wind that contains two sheet pinches
instead of two 
hot 
current sheets. Each pinch causes the magnetic field to
rotate through $\pi$ radians, so that the field outside of the pinches
is purely toroidal, and the magnitude of
the DC component is controlled by the location in phase of the
pinches. The analysis is particularly simple if the thickness of the
pinches is small --- i.e., they become
rotational discontinuities. This wave displays the
same radial evolution as the single, sinusoidal pinch
studied by \cite{kirkmochol11} (for details, see supplemental
material), the only difference is that the DC component
is $\propto1/r$, which causes the pinches to
migrate in phase. Ultimately, when the wave energy has been completely
converted into kinetic energy, the pinches merge. However, provided
their initial separation is not small, merging occurs only
towards the end of the acceleration phase, when the kinetic energy
flux is already comparable to the Poynting flux.

Pulsar winds are usually modeled as either
isotropic or axisymmetric, with the power concentrated towards the
equator \cite{porthetal17}, but the estimates presented below 
are not sensitive to this distinction. Assuming isotropy,
two parameters characterize the wind:
(i) the ratio at the light cylinder, $r_{\rm L}=cP/2\pi$,
of the gyrofrequency of a non-relativistic electron 
to the rotation frequency of the neutron star, given, 
for a magnetically dominated
flow, by
\begin{eqnarray}
a_{\rm L}&=&\left(e^2 L_{\rm s.d.}/m^2c^5\right)^{1/2}
\,=\,3.4\times10^{10}L_{38}^{1/2},
\end{eqnarray}
where $L_{\rm s.d.}=L_{38}\times 10^{38}\,\textrm{erg\,s}^{-1}$
is the spin-down power of the neutron star,
and (ii) the energy carried per particle in units of $mc^2$
\begin{eqnarray}
\mu&=&L_{\rm s.d.}/\left(\dot{N}_\pm mc^2\right),
\label{mudef}
\end{eqnarray}
where $\dot{N}_\pm$ is the rate at which electrons and positrons are
transported into the nebula by the wind.
For the Crab,
$a_{\rm L}\approx 7.6\times 10^{10}$, but $\mu$ is uncertain.
It can be related to the \lq\lq multiplicity\rq\rq\ parameter 
$\kappa$ used in modeling pair production near the pulsar
\cite{ceruttibeloborodov17} (conventionally, one sets
$\kappa=a_{\rm L}/\left(4\mu\right)$ \cite{lyubarskykirk01}), but
this should not be interpreted too literally. 
A latitude-dependent mass-loading of the wind can be 
defined by generalizing (\ref{mudef}), which enables detailed
modeling of the radio to X-ray emission
of the Crab Nebula \cite{olmietal15}. The tightest constraints, however, 
refer to the average of $\mu$
over the entire wind and over the lifetime of this object:
$10^4\lesssim\mu\lesssim10^6$, which corresponds to
$\dot{N}_\pm\approx 10^{39}$--$10^{41}\,\textrm{s}^{-1}$ and
$10^6\gtrsim\kappa\gtrsim10^4$\cite{bucciantinietal11}.

\begin{figure}
\includegraphics[width=8 cm]{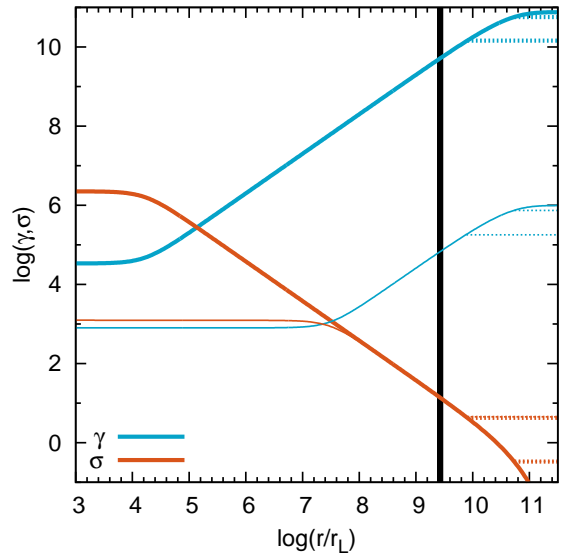}
\caption{\label{fig1}%
The radial evolution of the magnetization 
parameter 
$\sigma$ and fluid Lorentz factor $\gamma$ in a pulsar wind 
for high pair loading ($\mu=10^6$, thin lines) and low 
pair loading ($\mu=a_{\rm L}$, thick lines), for 
parameters corresponding to the Crab ($a_{\rm L}=7.6\times 10^{10}$). 
The position of the termination shock \cite{hesteretal02} is shown as a 
thick vertical line. 
For phase-averaged, DC~magnetic fields equal to 
$90\,\%$ and $50\,\%$ 
of the field magnitude at launch,
the horizontal, dotted lines show the solutions
after dissipation of the wave energy, 
i.e., in the regions $r\gtrsim10^{10}r_{\rm L}$ and $r\gtrsim10^{11}r_{\rm L}$, respectively. 
}
\end{figure}

The radial evolution of the wind is governed by three equations derived in 
Ref.~\cite{kirkmochol11} and in the 
supplemental material (where they are numbered (35), (37) and (38)). 
These can be integrated numerically
to find the three unknown functions
of $r$, which are the fluid Lorentz factor $\gamma$, 
the \lq\lq magnetization parameter\rq\rq\ 
$\sigma=\left|B\right|^2/\left(8\pi n_0\gamma^2 mc^2\right)$, with
$n_0$ the proper number density of the fluids, and 
$B$ the magnetic field strength, and 
the transverse component of the
dimensionless fluid four-velocity
$u_\bot=v_\bot/\left(c^2-v_\bot^2\right)^{1/2}$. 
Despite impressive progress based on 3D~MHD and force-free simulations
\cite{tchekhovskoyetal16}, the nature of the wind at launch, 
and, therefore, the initial conditions remain uncertain. 
Fundamental properties of
axisymmetric flows suggest they accelerate steadily up to the
sonic point, and thereafter coast at constant speed
(\cite{kirketal09}, see supplemental material).  Assuming 
the wave is launched as a mildly supersonic MHD flow, Fig.~\ref{fig1}
shows results for $a_{\rm L}=7.6\times 10^{10}$ and two different
values of mass-loading: $\mu=10^6$, the time-averaged value 
needed to provide the optical to X-ray
emitting particles in the Crab Nebula, and $\mu=a_{\rm
  L}$, a severely charge-depleted value with $\kappa\approx
1/4$. The latter is a plausible upper limit on $\mu$, since it
corresponds to the charge density at which light-like waves can
propagate already at radius $r_{\rm L}$. If these predominate,
the solutions shown in Fig.~\ref{fig1} lose validity. However,
embedded fluctuations could, in principle, be preserved at even lower
density --- 
a more rigorous but higher upper limit, $\mu\le a_{\rm
    L}^{3/2}$, is
  derived in equation (40) of the supplemental material. 
Note that a fluctuation in pair
loading need not be isotropic, and is advected radially with the flow.
Sectors of the wind separated by an angle $>1/\gamma$ evolve
essentially independently of each other, since a light signal launched
by a fluid element in one sector does not reach the equivalent element
in the other until it has more than doubled its radius.

The results shown in Fig.~\ref{fig1} exhibit three phases. 
In order of increasing radius, these
are (i) the MHD phase, in which $u_\bot$ is small, and both
$\gamma$ and $\sigma$ are constant; (ii) an
acceleration phase in which $u_\bot\sim 1$, and the inertia of the
charge carriers causes $\gamma$ to increase and $\sigma$ to decrease
at the expense of the oscillating component of the field; (iii) a
coasting phase in which the wave again proceeds at constant $\gamma$,
which begins either when the current sheets merge (for 
a finite DC component) or is approached asymptotically as
$\sigma\rightarrow0$.

An approximate solution for the acceleration phase 
is \cite{kirkmochol11}:
\begin{eqnarray}
u_\bot&\approx&1\qquad
\gamma\,\approx\,2\mu r/\left(a_{\rm L}r_{\rm L}\right)
\qquad \sigma\,\approx\,r_{\rm L}a_{\rm L}/\left(2r\right),
\label{accelerationzone}
\end{eqnarray}
valid for $a_{\rm L}\gamma_{\rm L}/\mu\ll r/r_{\rm L}\ll a_{\rm L}$, where
$\gamma_{\rm L}$ is the initial Lorentz factor of the wind. 
According to (\ref{accelerationzone}), 
inductive acceleration is a relatively slow process requiring an
undisturbed pulsar wind that extends to very large radius. For 
the Crab, Fig.~\ref{fig1} shows that $\sigma\approx 10$ at the
termination shock, independent of 
$\mu$. Thus, only $10\,\%$ of the Poynting flux is converted into
kinetic-energy flux before the wind reaches the
inner boundary of the Nebula.

Taking into account that 
$B\approx\left(2\pi m c/e P\right)\left(a_{\rm L}r_{\rm L}/r\right)$
for a magnetically dominated flow, and that the termination shock
compresses it by roughly a factor of three, the synchrotron emission of
an electron that enters the nebula with $\gamma$
given by Eq.~(\ref{accelerationzone}) peaks at a photon energy of
\begin{eqnarray}
h\nu_{\rm max}&\approx&18 \mu^2 \left(h/P\right)
\left[r_{\rm t.s.}/\left(a_{\rm L}r_{\rm L}\right)\right].
\label{numaxeq}
\end{eqnarray}
Thus, for the Crab under average conditions ($P=33\,\textrm{ms}$,
$r_{\rm t.s.}=4.3\times10^{17}\,\textrm{cm}$, $\mu=10^6$), particles
that cross the shock initially radiate at 
$h\nu_{\rm max}\approx 10^{-1}\,\textrm{eV}$, and are 
subsequently accelerated, either close to the shock 
or elsewhere in the Nebula, 
to produce the time-averaged optical to hard X-ray synchrotron emission.  
However, the situation changes dramatically if the
supply of particles in some section of the wind is interrupted. 
The arrival at the termination
shock of a low-density pocket with $\mu= a_{\rm L}$, 
as depicted in Fig.~\ref{fig1}, causes the injection
into the Nebula of a radially directed beam that initially radiates 
photons of energy $h\nu_{\rm max}\approx 500\,\textrm{MeV}$.

An electron injected at pitch angle $90^\circ$ into a homogeneous magnetic field with 
a Lorentz factor such that its synchrotron peak is $h\nu_{\rm max}$, 
is deflected through an angle 
\begin{eqnarray}
\delta\theta\left(\nu\right)&\approx&\left(80\,\textrm{MeV}/h\nu\right)
\left(1-\nu/\nu_{\rm max}\right)
\,\textrm{radians}
\label{deflectioneq}
\end{eqnarray}
whilst cooling to the point at which its peak emission 
has decreased to $h\nu$. Thus, if energetic
electrons are injected radially by a pulsar wind into the surrounding
nebula, synchrotron photons of energy $h\nu >
80\,\textrm{MeV}$ will appear to an observer with sufficient angular
resolution to emerge from a finite-sized patch on the termination
shock, centered on the pulsar. The precise size and shape of this patch
depend on the configuration of the magnetic field downstream of the 
termination shock, which is expected to be turbulent on length 
scales of $r_{\rm t.s.}$ \cite{porthetal16}, but, nevertheless, 
predominantly toroidal. A rough upper limit on the 
area of the patch, 
$A\lesssim\delta\theta^2\left(\nu\right)r_{\rm t.s.}^2$, follows from 
assuming random deflections through an angle of at
most that given in Eq.~(\ref{deflectioneq}). For $\nu\ll\nu_{\rm max}$, this implies 
$A\propto\nu^{-2}$. However, a weaker dependence on frequency is found 
if the beam diverges diffusively. 

\begin{figure}
\includegraphics[width=8 cm]{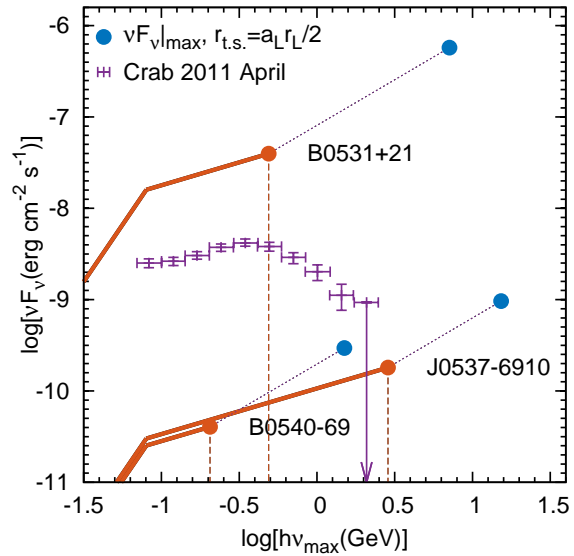}
\caption{\label{fig2}%
The predicted flare spectrum (Eq.~(\ref{predictedspectrum}), solid lines), for the 
three most powerful known pulsars: the Crab 
(B0531$+$21), and two objects in the Large Magellanic Cloud, assuming 
a turnover at $h\nu_{\rm t}=80\,\textrm{MeV}$ 
and a filling factor $f=1$.  
Dotted lines trace the locus of  
the peak flux as the position of the termination shock is varied between 
the observed value (orange dots)  and $a_{\rm L}r_{\rm L}/2$ 
(blue dots (circles in black and white)). 
Fermi observations of the powerful flare from the Crab Nebula 
in April~2011 are also shown --- points taken from Fig~6, epoch~7 of Ref.~\cite{buehleretal12}.
}
\end{figure}

This has important implications for the spectrum, time-dependence and
overall power of the emission: Assume the pulsar wind is depleted of
charges in a cone that occupies a solid angle $\Omega$ and includes
the line-of-sight to the observer. Then, photons with energy $h\nu >
h\nu_{\rm t}\approx\left(80/\Omega^{1/2}\right)\,\textrm{MeV}$, that are radiated by 
electrons crossing the termination shock in this cone, 
remain within it, and their steady-state, differential flux 
has the form typical of mono-energetically injected,
cooling electrons.  On the other hand, lower frequency photons emerge
from a patch with $A>\Omega r_{\rm t.s.}^2$, and are
radiated into a solid angle that exceeds $\Omega$. This depletes the
observed flux, leading to a  turnover at $\nu_{\rm
  t}$.  Using (\ref{deflectioneq})
to estimate the rate at 
which the beam of cooling electrons diverges, 
and employing a crude, monochromatic approximation for the
synchrotron emissivity, gives, for the differential energy flux
$F_\nu$:
\begin{eqnarray}
\nu F_\nu&=&\left\lbrace
\begin{array}{l}
f L_{\rm s.d.}/\left(8\pi\sigma D^2\right)
\left(\nu/\nu_{\rm max}\right)^{1/2} \\
\qquad \textrm{for }\nu_{\rm t}<\nu<\nu_{\rm max}\\
f L_{\rm s.d.}/\left(8\pi\sigma D^2\right)
\left(\nu_{\rm t}/\nu_{\rm max}\right)^{1/2}
\left(\nu/\nu_{\rm t}\right)^{5/2}\\
\qquad \textrm{for }\nu<\nu_{\rm t}
\end{array}
\right.
\label{predictedspectrum}
\end{eqnarray}
where $D$ is the distance to the source,
and the filling factor, 
$f$, ($\le1$) is the fraction of the 
flow containing charge-depleted regions.
As shown in Fig.~\ref{fig2},
observations of powerful gamma-ray flares from the Crab Nebula show a
spectral form roughly consistent with (\ref{predictedspectrum}),
provided $h\nu_{\rm max}$ given by Eq.~(\ref{numaxeq}) is 
$500\,\textrm{MeV}$, and $\Omega$ is about $1\,\textrm{sr}$.
Furthermore, 
the available power is more than adequate to explain the 
most powerful flare observed to date (that of April 2011\cite{buehleretal12}):
within the uncertainties in the distance to this object,
and the angular distribution of power in the pulsar wind, 
this flare is consistent with $f\approx 0.1$
and $\sigma\approx10$ at the termination
shock, as indicated in Fig.~\ref{fig1}.
Gamma-ray flares from the Crab are seen with a range of powers and
peak photon energies, and collations of their spectra, as presented,
for example, in Figs.~6 and 7 of Ref.~\cite{buehleretal12}, can, in
principle, be been used to test model predictions. However, although
the crude model of synchrotron radiation plotted in Fig.~\ref{fig2}
suggests rough agreement with the data, a more sophisticated approach
would be needed to fit details of the spectrum. 

Since electron-positron pairs are injected into the wind relatively
close to the pulsar, the relevant timescale of variations is roughly the
rotation period. A reduction in the injection rate on this timescale causes the wind
to propagate at a higher Lorentz factor. At the termination shock,
the transition between the two flow states is broadened in time to roughly
$r_{\rm t.s.}/c\gamma^2$, where $\gamma$ refers to the slower flow,
but this is still very short 
($\sim 10\,$s for the Crab) 
compared to the variation time of the
flares. Therefore, both the leading edge and the trailing edge of a
cone of charge-depleted wind can be treated as sharp transitions, and the observed
rise and decay-times of the emission are dominated by the
variation in travel time from different parts of the illuminated patch
on the termination shock. This patch appears larger at lower photon
energy, and the timescale $t_{\rm var}$ of variation is,
correspondingly, energy dependent:
\begin{eqnarray}
t_{\rm var}&=& \delta\theta^2\left(\nu\right)r_{\rm t.s.}/c\\
&\approx&167\, 
\left(80\,\textrm{MeV}/h\nu\right)^2
\left(1-\nu/\nu_{\rm max}\right)^2 
\textrm{days}
\nonumber
\end{eqnarray}
where, in the second equation, the value of $r_{\rm t.s.}$ for the
Crab has been inserted. Thus, the timescale of eight hours observed in
the powerful 2011~April Crab flare \cite{buehleretal12} suggests a
patch of angular extent $\delta\theta\approx 2.5^\circ$. This is
consistent with $h\nu\approx 400\,\textrm{MeV}$, provided
$h\nu_{\rm max}\approx 500\,\textrm{MeV}$, which is also suggested by
the spectrum shown in Fig.~\ref{fig2}. 

Although it provides an attractive scenario, the main shortcoming of
the theory presented above is that it does not explain 
why the supply of charged particles to small parts of a
magnetically dominated, relativistic outflow should suffer
interruptions.  The reason is assumed to lie in the physics of
the electromagnetic cascades that are responsible for producing the
pairs we observe when the flows terminate.  These are known to be
non-stationary, both in pulsars and in black hole magnetospheres
\cite{ceruttibeloborodov17,levinsonsegev17}, but a detailed
understanding of their 3D spatial and temporal properties is not
currently within reach. On the other hand, as long as the theory
provides a reasonable fit to the data, it can be used to infer the
properties of these cascades. Indeed, the fact that flares from the
Crab can last for several days or longer implies that 
during this time, pockets of depleted charge in 
a small cone of outflow directed at the observer
have a spatial or temporal filling factor of several per cent, 
which, in turn, suggests that the global
geometry of the cascade varies on a timescale much longer than the
pulsar rotation period. 

The Crab is the most powerful pulsar known in the Milky Way, but it is
not unique. In our neighboring galaxy, the Large Magellanic Cloud, two
pulsars of comparable power are known: PSR~J0537$-$6910 and
PSR~B0540$-$69 \cite{Fermi_LMC_pulsar15}. The predicted flare spectra
for these objects are shown in Fig.~\ref{fig2}, using estimates of the
location of the termination shock given in Table~2 of reference
\cite{kargaltsevpavlov08}. Flares from the gamma-ray pulsar B0540$-$69
are not expected at energies significantly larger than
$200\,\textrm{MeV}$, and may, therefore, be difficult to detect. 
PSR~J0537$-$6910, however, has a shorter pulse period and a
longer undisturbed wind than the Crab.  As a
result, its nebula (PWN N~157B) should exhibit
synchrotron flares with photons of up to $3\,\textrm{GeV}$, as
suggested by a recent analysis of Fermi-data~\cite{saitoetal17}. Thus,
inductive spikes may be a common property of young, powerful pulsars,
and enable more detailed modeling and a deeper understanding of the
way in which they energize their surrounding nebulae. 
Furthermore, inasmuch as they
are also powered by low density, magnetically dominated, relativistic
outflows, both blazars and gamma-ray bursts may exhibit analogous 
phenomena~\cite{kirkmochol11}. 
 
\begin{acknowledgments}
  This research was supported by a Grant from the GIF, the
  German-Israeli Foundation for Scientific Research and Development.
\end{acknowledgments}
%

\end{document}